\documentclass[aps,prl,reprint,amsmath,amsfonts,amssymb,superscriptaddress]{revtex4-1}
\usepackage{graphicx,calc}
\usepackage{color,bm}
\usepackage{ulem}
\usepackage{braket}
\usepackage[colorlinks,urlcolor=blue,citecolor=blue,linkcolor=blue]{hyperref}

\begin{document}
	
\title{Efficient sorting of orbital-angular-momentum states with large topological charges and their unknown superpositions via machine learning}

\author{Ling-Feng Zhang}
\thanks{These authors contributed equally to this work.}
\affiliation{Guangdong Provincial Key Laboratory of Quantum Engineering and Quantum Materials, GPETR Center for Quantum Precision Measurement and SPTE, South China Normal University, Guangzhou 510006, China}
\author{Ya-Yi Lin}
\thanks{These authors contributed equally to this work.}
\affiliation{Guangdong Provincial Key Laboratory of Quantum Engineering and Quantum Materials, GPETR Center for Quantum Precision Measurement and SPTE, South China Normal University, Guangzhou 510006, China}
\author{Zhen-Yue She}
\affiliation{Guangdong Provincial Key Laboratory of Quantum Engineering and Quantum Materials, GPETR Center for Quantum Precision Measurement and SPTE, South China Normal University, Guangzhou 510006, China}

\author{Zhi-Hao Huang}
\affiliation{Guangdong Provincial Key Laboratory of Quantum Engineering and Quantum Materials, GPETR Center for Quantum Precision Measurement and SPTE, South China Normal University, Guangzhou 510006, China}
\author{Jia-Zhen Li}
\affiliation{Guangdong Provincial Key Laboratory of Quantum Engineering and Quantum Materials, GPETR Center for Quantum Precision Measurement and SPTE, South China Normal University, Guangzhou 510006, China}
\author{Hui Yan}
\affiliation{Guangdong Provincial Key Laboratory of Quantum Engineering and Quantum Materials, GPETR Center for Quantum Precision Measurement and SPTE, South China Normal University, Guangzhou 510006, China}
\affiliation{Frontier Research Institute for Physics, South China Normal University, Guangzhou 510006, China}
\author{Wei Huang}\email{WeiHuang@m.scnu.edu.cn}
\affiliation{Guangdong Provincial Key Laboratory of Quantum Engineering and Quantum Materials, GPETR Center for Quantum Precision Measurement and SPTE, South China Normal University, Guangzhou 510006, China}
\author{Dan-Wei Zhang}\email{danweizhang@m.scnu.edu.cn}
\affiliation{Guangdong Provincial Key Laboratory of Quantum Engineering and Quantum Materials, GPETR Center for Quantum Precision Measurement and SPTE, South China Normal University, Guangzhou 510006, China}
\affiliation{Frontier Research Institute for Physics, South China Normal University, Guangzhou 510006, China}
\author{Shi-Liang Zhu}
\affiliation{Guangdong Provincial Key Laboratory of Quantum Engineering and Quantum Materials, GPETR Center for Quantum Precision Measurement and SPTE, South China Normal University, Guangzhou 510006, China}
\affiliation{Frontier Research Institute for Physics, South China Normal University, Guangzhou 510006, China}

\date{\today}

\begin{abstract}
Light beams carrying orbital-angular-momentum (OAM) play an important role in optical manipulation and communication owing to their unbounded state space. However, it is still challenging to efficiently discriminate OAM modes with large topological charges and thus only a small part of the OAM states have been usually used. Here we demonstrate that neural networks can be trained to sort OAM modes with large topological charges and unknown superpositions. Using intensity images of OAM modes generalized in simulations and experiments as the input data, we illustrate that our neural network has great generalization power to recognize OAM modes of large topological charges beyond training areas with high accuracy. Moreover, the trained neural network can correctly classify and predict arbitrary superpositions of two OAM modes with random topological charges. Our machine learning approach only requires a small portion of experimental samples and significantly reduces the cost in experiments, which paves the way to study the OAM physics and increase the state space of OAM beams in practical applications.

\end{abstract}

\maketitle

{\color{blue}\textit{Introduction.---}}The spatial modes of light, such as the Laguerre-Gaussian (LG) modes, give access to an in principle unbounded state space for encoding information beyond one bit per photon ~\cite{PhysRevLett.75.826,Dholakia2011,Simpson:97,Grier2003,Padgett2011,Gibson:04,Wang2012,Bozinovic1545,Krenn13648,Laverye1700552,Mair2001,PhysRevLett.88.013601,PhysRevLett.89.240401,PhysRevLett.90.133901,Leach662,Dada2011,Wang2015}. LG modes carry well-defined $\ell\hbar$ quanta of orbital-angular momentum (OAM), with arbitrary integer $\ell$ as the topological charge. Owing to the infinite dimensions of $\ell$, OAM beams offer an ideal source both in classical and quantum optics with
various applications, such as optical manipulation~\cite{PhysRevLett.75.826,Dholakia2011,Simpson:97,Grier2003,Padgett2011}, communications~\cite{Gibson:04,Wang2012,Bozinovic1545,Krenn13648,Laverye1700552}, high-dimensional quantum information~\cite{Mair2001,PhysRevLett.88.013601,PhysRevLett.89.240401,PhysRevLett.90.133901,Leach662,Dada2011,Wang2015} and so on. In these applications, separating and discriminating the OAM eigenstates play an important role, but are still challenging. Several techniques have been developed to overcome this difficulty, such as sophisticated hologram designs~\cite{PhysRevLett.96.163905,Wen:20}, log-polar coordinate transformation methods~\cite{PhysRevLett.105.153601,Mirhosseini2013}, utilizing an optical resonantor~\cite{Wei2020} or a photodetector with carefully fabricated electrode geometries~\cite{Ji763}. Nevertheless, because of the lack of technologies to separate and discriminate OAM states with large topological charges, only a small part of the OAM states are used in practical applications.

Recently, owing to the ability to classify, identify and interpret massive data, machine learning (ML) has been applied in different areas of physics \cite{Carleo2019}. The ML methods based on neural networks are emerging as a versatile toolbox to tackle a variety of hard tasks arising in experimental platforms~\cite{PhysRevLett.122.210503,Rem2019,PhysRevLett.125.127401,PhysRevLett.126.060401,ai2021experimentally,guo2021machine,hofer2020atom,lode2020optimized,Rodrigues2021,Sivaraman2021,LFZhang2021}. Remarkably, the convolutional neural network (CNN) schemes have been proposed to detect non-OAM and misaligned (distorted) OAM beams \cite{Hofer2019,Zhao2018,Bhusal2021} and to demultiplex OAM modes under turbulence \cite{Doster2017}. Recent applications of CNNs have focused on distinguishing OAM modes with fractional topological charges ~\cite{PhysRevLett.123.183902,Na2021} and classifying and extracting the coefficients of given OAM superpositions~\cite{PhysRevLett.124.160401,pinheiro2020machine}, wherein only OAM modes of small topological charges ($|\ell|\leq5$) and given superposition states are recognized. Hitherto, due to the fact that only OAM modes with small or given $\ell$ are experimentally generated and used for CNNs \cite{PhysRevLett.123.183902,Na2021,PhysRevLett.124.160401,pinheiro2020machine}, the power of ML in efficient sorting of OAM modes with large topological charges and unknown superpositions has not yet been demonstrated in laboratory.

In this Letter, we demonstrate that the CNN can be trained to efficiently recognize OAM modes with large topological charges and their unknown superpositions, even with a small portion of the experimentally generated data.
Using intensity profiles of OAM modes generalized in simulations and experiments as the input data, based on the regression and classification analysis, we illustrate that our neural network has great generalization power to recognize OAM modes of large topological charges ($|\ell|$ up to 25) beyond training areas with nearly $100\%$ accuracy. Remarkably, our trained CNN can correctly classify and predict distinct and arbitrary superpositions of two OAM modes with random topological charges. Since our approach only requires a small portion of experimental samples, it significantly reduces the cost in experiments. Our results showcase the potential of the ML approach to study the OAM physics and increase the state space of the OAM modes in practical applications.

\begin{figure*}[!htb]
  \includegraphics[width=0.8\textwidth]{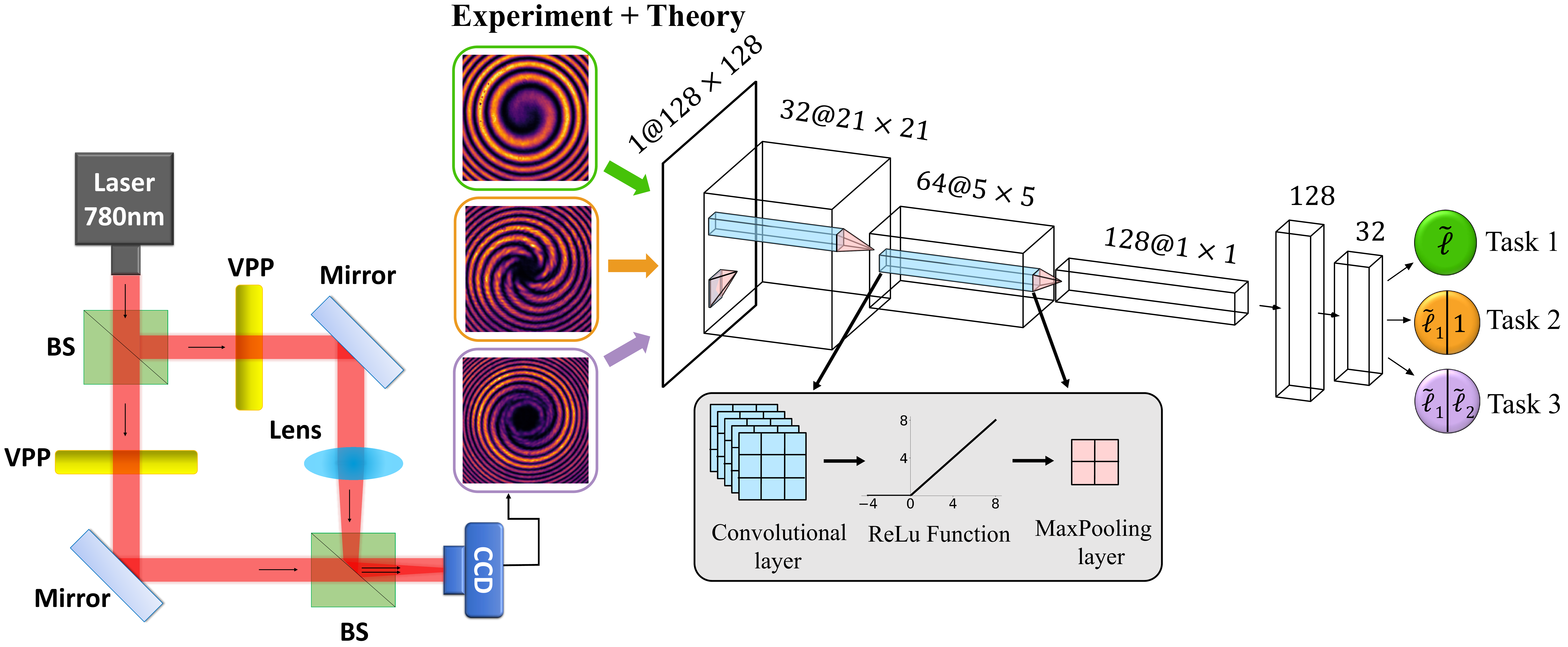}
  \caption{(Color online) Schematic of the experimental setup and the ML workflow. The linearly polarized laser beam of 780nm wavelength goes through a non-polarizing beam splitter (BS) and is divided into two parts. The Gaussian modes change into the LG modes with OAMs. One of the OAM beams is propagated through a lens and its wavefront transforms into a spherical wave. The vortex phase plates (VPPs) are added to create tunable OAMs. Then two OAM modes are combined together using another BS, and the interference pattern is recorded by a CCD camera. The measured and simulated intensity profiles are input to the CNN. The CNN consist of several blocks including convolutional layer, nonlinear activation function, max-pooling layer and fully connected layer \cite{SM}, where the number and size of each part are indicated. For Task 1, the CNN is trained to predict the topological charge of a LG mode as a regression analysis. For Tasks 2 and 3, the CNN are trained to predict the topological charges of superpositions of two LG modes as a classification analysis.}
   \label{fig1}
\end{figure*}

{\color{blue}\textit{Experimental generation of OAM modes.---}}
LG modes are the solutions of the free space Maxwell's equations within the paraxial approximation. For a beam propagating along the $z$ axis, the LG modes can be described by the LG functions in cylindric coordinates $\text{LG}_{\ell,p}(r,\theta,z)=N(r)L^{|\ell|}_{p}(\frac{2r^2}{w_{z}^{2}})
    \exp{[-\frac{r^2}{w_{z}^{2}}+i(\frac{kr^2}{2R_z}-kz + \ell\theta-m\phi_g)]}$, with $N(r)=\sqrt{\frac{2p!}{\pi(p+|\ell|)!}}\frac{1}{w_z}(\frac{\sqrt{2}r}{w_z})^{|\ell|}$.
Here $\ell$ is the topological charge of the OAM modes, $p$ is the radial mode number, $L^{|\ell|}_{p}$ are the Laguerre polynomials, the beam parameters $w_z=w_0\sqrt{1+(\frac{z}{z_R})^2}$ denotes the beam waist with $w_0$ being the waist at the focus and $z_R=\frac{\pi w_{0}^{2}}{\lambda}$ being the Rayleigh range, $R_z=z[1+(\frac{z}{z_R})^2]$ is the radius of curvature, $k=\frac{2\pi}{\lambda}$ is the wave number with $\lambda$ being the wavelength, and $\phi_g=\arctan{(\frac{z}{z_R})}$ denotes the Gouy phase and is multiplied with the mode order $m=(2p+|\ell|+1)$.

Hereafter we focus on the topological charge $\ell$ and ignore the radial number by setting $p=0$ for the purpose of this work. Then the LG functions can be rewritten as $\text{LG}_{\ell,0}\doteq\text{LG}_{\ell}$ for simplicity. We consider a single LG mode and superpositions of two OAM modes, which are denoted by $\ket{\psi_{\ell}}=\ket{\text{LG}_{\ell}}$ and $\ket{\psi_{\ell_1,\ell_2}} = \cos\frac{\theta}{2}\ket{\text{LG}_{\ell_1}} + e^{i\phi}\sin\frac{\theta}{2}\ket{\text{LG}_{\ell_2}})$ with $\phi\in[0,2\pi]$ and $\theta\in(0,\pi)$, respectively. The corresponding intensity distributions (images) in the $x$-$y$ place (perpendicular to the propagating direction) $\text{I}_{\ell}=|\langle\psi_{\ell}\ket{\psi_{\ell}}|^2$ and $\text{I}_{\ell_1,\ell_2}=|\langle\psi_{\ell_1,\ell_2}\ket{\psi_{\ell_1,\ell_2}}|^2$ are taken as the input data for training and testing the CNNs, which are both generated in numerical simulations and experiments. As shown in Fig.~\ref{fig1}, in the first task, we aim to sort and predict large topological charges $\ell$ of the LG mode $\ket{\psi_{\ell}}$ after training the CNN. In the second and third tasks, we intend to classify the superposition of two OAM modes $\ket{\psi_{\ell_1,\ell_2}}$ and predict the unknown (random) topological charges $\ell_1$ and $\ell_2$. We use images with different $\phi$, concentrate on the equivalent superposition with $\theta=0.5\pi$ first, and finally generalize our results to arbitrary superpositions with various $\theta$. Thus, our ML approach not only recognizes OAM modes with different superposition coefficients \cite{PhysRevLett.124.160401,pinheiro2020machine}, but also simultaneously predicts their random and large topological charges.

The experimental setup for generating OAM modes is shown in Fig.~\ref{fig1}. The linearly polarized laser beam of wavelength 780nm goes through a non-polarizing beam splitter  and then is divided into two beams. The vortex phase plates are applied to create the LG modes with tunable topological charges $\ket{\ell_1}$ and $\ket{\ell_2}$ of the spherical and plane wave OAMs~\cite{doi:10.1119/1.18283}, with $\ket{\ell_{1,2}}$ up to $8$ accessible in our experiments. One beam propagates through a lens ($f=300$mm) and its wavefront transforms as a spherical wave. Then the two OAM beams are combined together using the second beam splitter, which leads to the superposition state $\ket{\psi_{\ell_1,\ell_2}}$. The interference pattern is finally recorded by a CCD camera ($1294\times960$ pixels, 3.75$\mu m$ $\times$ 3.75$\mu m$ per pixel) and down sampled to $128\times128$ pixels as the input data for the CNN. By setting $\ell_2=0$ and $\ell_1=\ell$ through the vortex phase plates, we obtain the intensity pattern of the LG mode $\ket{\psi_{\ell}}$ in this case. For given $\ell$ or $\ell_{1,2}$, we experimentally create hundreds of intensity images with different (rotation or $\phi$) phases.

\begin{figure}
    \centering
    \includegraphics[width=0.48\textwidth]{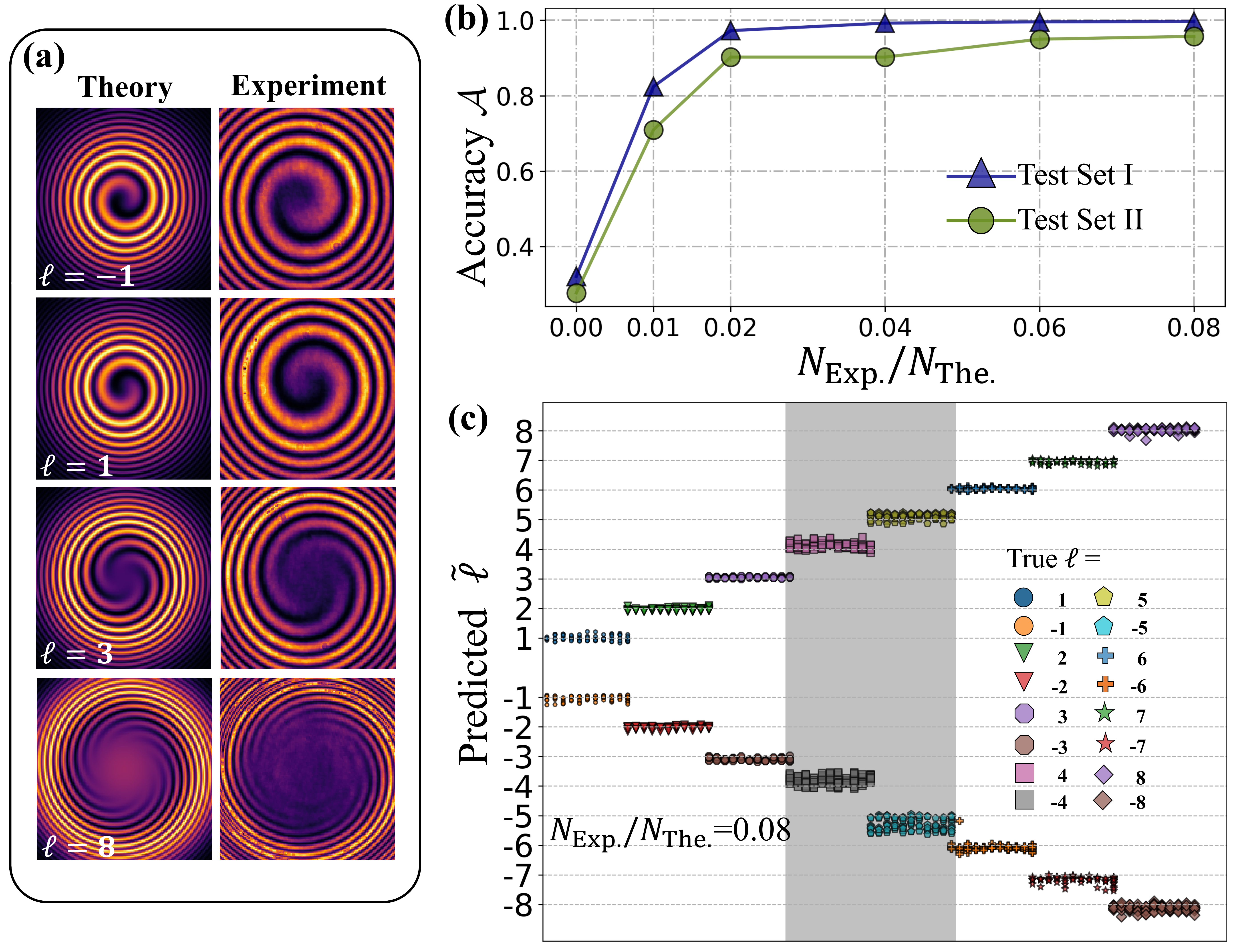}
    \caption{(Color online) (a) Theoretical (left) and experimental (right) images show the intensity distributions of $\ket{\psi_{\ell}}$. The number of spiral fringes is equal to the topological charge $|\ell|$, and the spiral direction is clockwise if $\ell>0$ or counterclockwise otherwise. (b) The average accuracy $\mathcal{A}$ for two test sets against the fraction of experimental images $N_{\text{Exp.}}/N_{\text{The.}}$. (c) The predicted topological charge $\tilde{\ell}$ at $N_{\text{Exp.}}/N_{\text{The.}}=0.08$. The unseen intervals $\{\pm4,\pm5\}$ during the training are marked as the shaded areas.}
    \label{fig2}
\end{figure}

{\color{blue}\textit{Sorting multi-OAM modes.---}}We first demonstrate that the trained CNN can retrieve the topological charges of the OAM beam in Task 1. In this case, we do the training as a regression analysis, which endows the CNN to recognize the unseen data areas. We consider the OAM modes $\ket{\psi_{\ell}}$ with true topological charges $\ell \in\{\pm1,\pm2,\pm3,...,\pm8\}$, with several typical intensity images shown in Fig.~\ref{fig2}(a). The intensity images consist of spiral fringes, which arise from the combination of the radial phase variation due to wave-front curvature and the tangential phase variation due to the azimuthal phase dependence. The number of spiral fringes is equal to $|\ell|$, and the spiral direction is clockwise if $\ell>0$ or counterclockwise otherwise.

The architecture of our CNN is shown in Fig.~\ref{fig1}. We define a block including convolutional layer, rectified linear unit (Relu) as nonlinear activation function and max-pooling layer. The CNN has three blocks. We use max-pooling layers with the kernel size of $2\times2$ with 2 strides for all blocks. The convolutional layers of three blocks have 32 kernels of size $5\times5$ with 3 strides, 64 kernels of size $3\times3$ with 2 strides and 128 kernels of size $3\times3$ with 2 strides, respectively, followed by a fully-connected layer with 32 neurons before output layer. The CNN is first trained with numerically generated data [e.g., the intensity images at the left row of Fig. \ref{fig2}(a)] for the true topological charge $\ell\in \{\pm1,\pm2,\pm3,\pm6,\pm7,\pm8\}$. For each $\ell$, we generate 400 images with different rotation phases, which can meanwhile improve the rotary robustness of our CNN. Thus, the entire training set contains $N_{\text{Total}}=4800$ images labeled by 12 different $\ell$s. Then we incrementally add experimentally generated data [e.g., the intensity images at the right row of Fig. \ref{fig2}(a)] with the number $N_{\text{Exp.}}$, while maintaining the data set size $N_{\text{Total}}=N_{\text{Exp.}}+N_{\text{The.}}=4800$ with $N_{\text{The.}}$ being the number of theoretical images. The weight parameters of the CNN are trained by using the Adam optimizer with batch size 128 for 100 epochs, the initial learning rate is set to be 0.01, and a $\mathcal{L}2$ regularization is used with weight decay of 0.0001 to prevent over-fitting \cite{paszke2019pytorch,LeCun2015,Shorten2019}. The output of the trained CNN is a real number $\tilde{\ell}$, which is close to an integer and can be interpreted as the predicted topological charge. We determine the percentage of the correct predictions as the accuracy $\mathcal{A}$ by comparing the predicted topological charge $\tilde{\ell}$ with the true value $\ell$ ~\cite{SM}.

After the training, we test the CNN with two test sets with 100 experimental images per topological charge class for: (i) $\ell\in\{\pm1,\pm2,\pm3,\pm6,\pm7,\pm8\}$ as Test set I; and (ii) $\ell\in\{\pm4,\pm5\}$ as Test set II unseen by the CNN during training. The test results are presented in Figs.~\ref{fig2}(b-c). Figure~\ref{fig2}(b) shows the average accuracy $\mathcal{A}$ of two test sets against the fraction of experimental images added to the training set $N_{\text{Exp.}}/N_{\text{The.}}$. Note that the added experimental images are synthesized with avoiding intervals $\ell\in\{\pm4,\pm5\}$. We obtain two interesting results: i) The accuracy for both Test sets I and II increases as the experimental data added to the training data set. The mean values of predicted $\tilde{\ell}$ are closer to true $\ell$ and their standard deviations decrease with increasing of $N_{\text{Exp.}}/N_{\text{The.}}$ (see Fig. \ref{figS2} in the SM \cite{SM}). This is due to the fact that the addition of experimental images to the training set improves the capability of the CNN when taking into account deviations of the experimental data. Notably, a high accuracy $\mathcal{A}>0.95$ is already obtained when the small fraction of the training set is composed of experimental images $N_{\text{Exp.}}/N_{\text{The.}}=6\%$. ii) We find that due to generality of the trained CNN, it is capable of predicting the unseen intervals with a high accuracy. The predicted topological charges $\tilde{\ell}$ for the two test sets when $N_{\text{Exp.}}/N_{\text{The.}}=8\%$ are given in Fig.~\ref{fig2}(c), showing the distribution of $\tilde{\ell}$s close to the true values $\ell$s.

\begin{figure}
    \centering
    \includegraphics[width=0.4\textwidth]{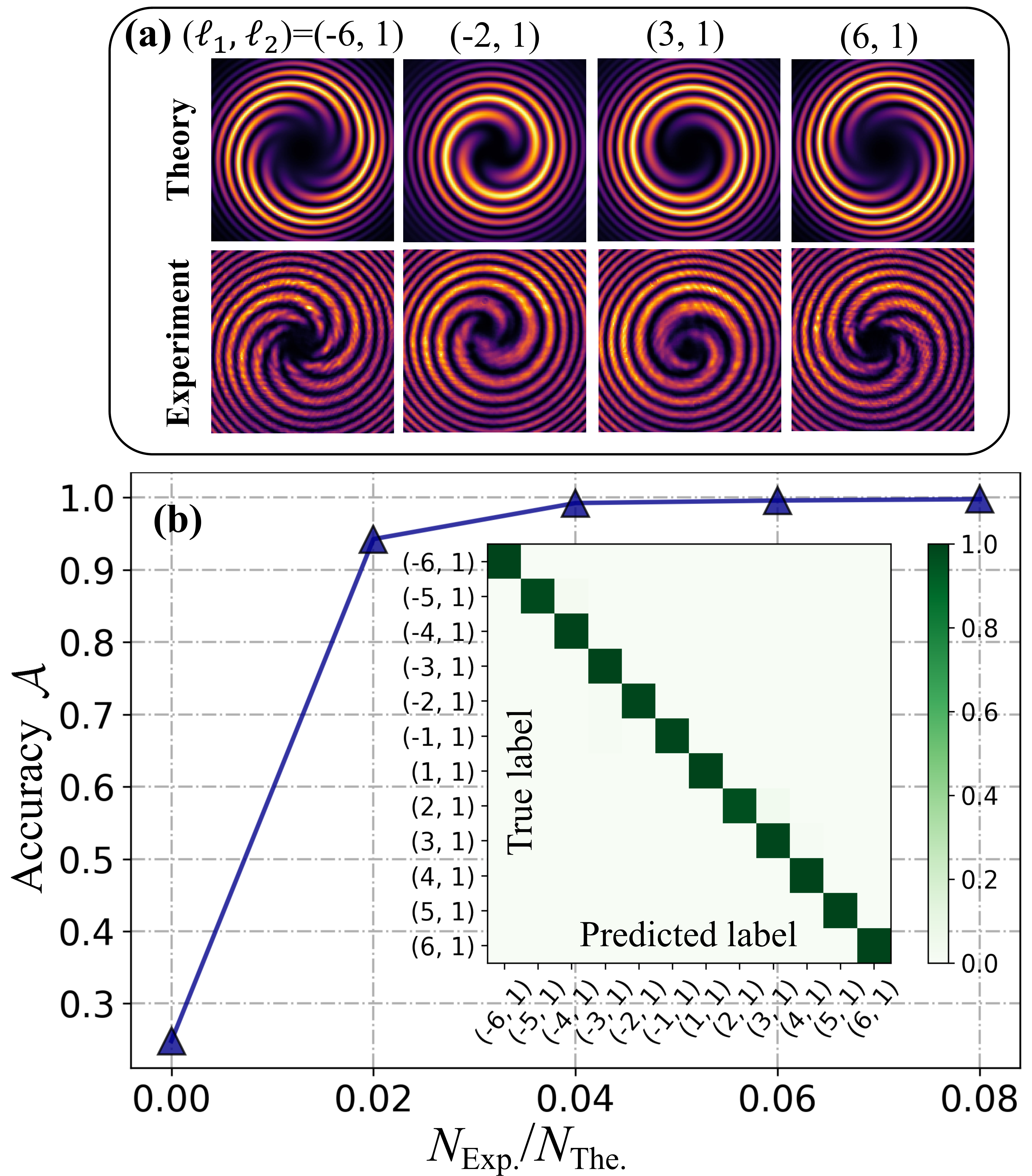}
    \caption{(Color online) (a) Theoretical (up) and experimental (down) images show the intensity distributions of the superposition states $\ket{\psi_{\ell_1,1}}$. (b) Average accuracy $\mathcal{A}$ of the test set against the fraction of experimental images $N_{\text{Exp.}}/N_{\text{The.}}$. The inset shows the confusion matrix predicted by the CNN trained at $N_{\text{Exp.}}/N_{\text{The.}}=0.04$.}
    \label{fig3}
\end{figure}

{\color{blue}\textit{Classifying unknown superpositions of OAM modes.---}}We consider two following tasks of obtaining the topological charges of superpositions of two OAM modes $\ket{\psi_{\ell_1,\ell_2}}$. In Task 2, we consider the superpositions of a random OAM mode with $\ell_1\in\{\pm1,\pm2,\pm3,\pm4,\pm5,\pm6\}$ and a fixed OAM mode with $\ell_2=1$, which are denoted by $\ket{\psi_{\ell_1,1}}$. Several intensity images in this task are shown in Fig.~\ref{fig3}(a). The architecture of the CNN is based on previous regression analysis, but the output layer is replaced by 12 neurons here as a classification analysis. The output of the CNN is the normalized probabilities corresponding to 12 classes of different values of true topological charges $\ell_1$. The predicted topological charge $\tilde{\ell}_1$ is interpreted as the class that has the maximum probability. The accuracy $\mathcal{A}$ in this task refers to the fraction of correctly classified images. The size of the training set is 400 images per class. We collect 100 experimental images per class as the test set. The training procedure is the same as that for Task 1. After the training, we use the test set to evaluate the performance of the CNN. The average accuracy of the test set increases significantly with increasing the fraction of experimental images added to the training set, as shown in Fig.~\ref{fig3}(b). Remarkably, the trained CNN can successfully identify all random superposition states $\ket{\psi_{\ell_1,1}}$ and predict the topological charges $\ell_1$, with a very high accuracy $\mathcal{A}\approx99\%$ even under a small portion of the experimental data $N_{\text{Exp.}}/N_{\text{The.}}=4\%$, as shown in the inset of Fig. \ref{fig3}(b).

\begin{figure}
    \centering
    \includegraphics[width=0.4\textwidth]{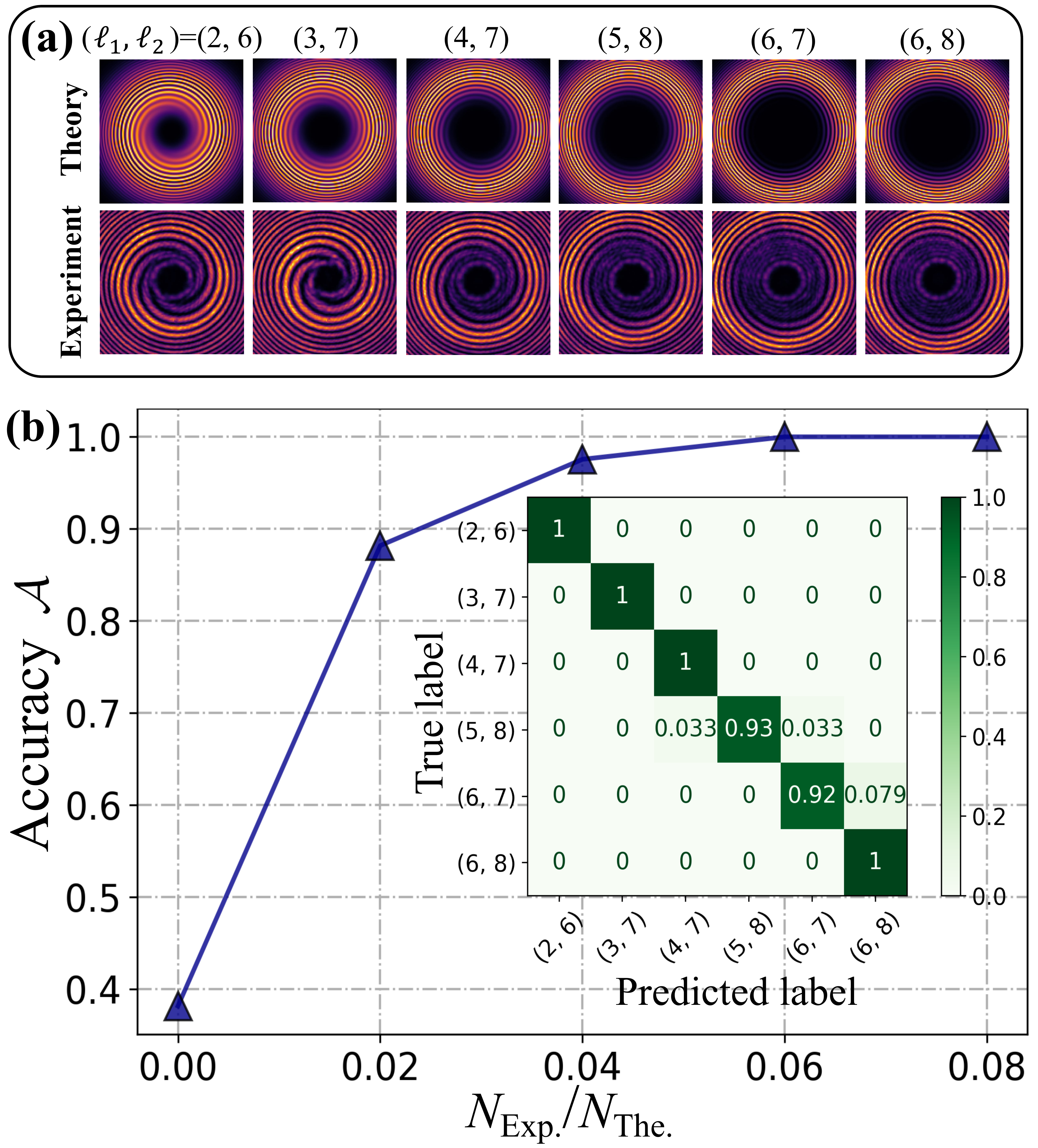}
    \caption{(Color online) (a) Theoretical (up) and experimental (down) images show the intensity distributions of the superposition states $\ket{\psi_{\ell_1,\ell_2}}$.
    (b) Average accuracy $\mathcal{A}$ against the fraction $N_{\text{Exp.}}/N_{\text{The.}}$. The inset shows the confusion matrix predicted by the CNN trained at $N_{\text{Exp.}}/N_{\text{The.}}=0.04$.}
    \label{fig4}
\end{figure}

We consider the superpositions of two random OAM modes $\ket{\psi_{\ell_1,\ell_2}}$ as a more general situation in Task 3. For the sake of the computational capability and visualized presentation, we select the following 6 classes $\{\ket{\psi_{2,6}}, \ket{\psi_{3,7}},\ket{\psi_{4,7}},\ket{\psi_{5,8}},\ket{\psi_{6,7}},\ket{\psi_{6,8}}\}$, as shown in Fig.~\ref{fig4}(a).
Note that the spiral number and direction of $\ket{\psi_{2,6}}$ ($\ket{\psi_{4,7}}$) are the same as those of $\ket{\psi_{3,7}}$ ($\ket{\psi_{5,8}}$). Their intensity distributions are just slightly different in the radial direction, which are hard to discriminate for human eyes. Extraordinarily, we show that the CNN can precisely detect these small differences. The workflow in this task is similar to the previous cases, except that the output layer of the CNN now has 6 neurons. The average accuracy of the test set against the fraction of experimental images added to the training set as shown in Fig.~\ref{fig4}(b). We obtain two interesting results: i) the accuracy of the trained CNN increases as the experimental data added to the training data set. Notably, we only need almost $4\%$ of the training set composed of experimental images, then CNN can perform great on experimental cases (with a probability of $99\%$). ii) The CNN can discriminate between the superposition states $\ket{\psi_{2,6}}$ ($\ket{\psi_{4,7}}$) and $\ket{\psi_{3,7}}$ ($\ket{\psi_{5,8}}$) with high accuracy, as shown in the inset of Fig.~\ref{fig4}(b).

\begin{figure}
    \centering
    \includegraphics[width=0.4\textwidth]{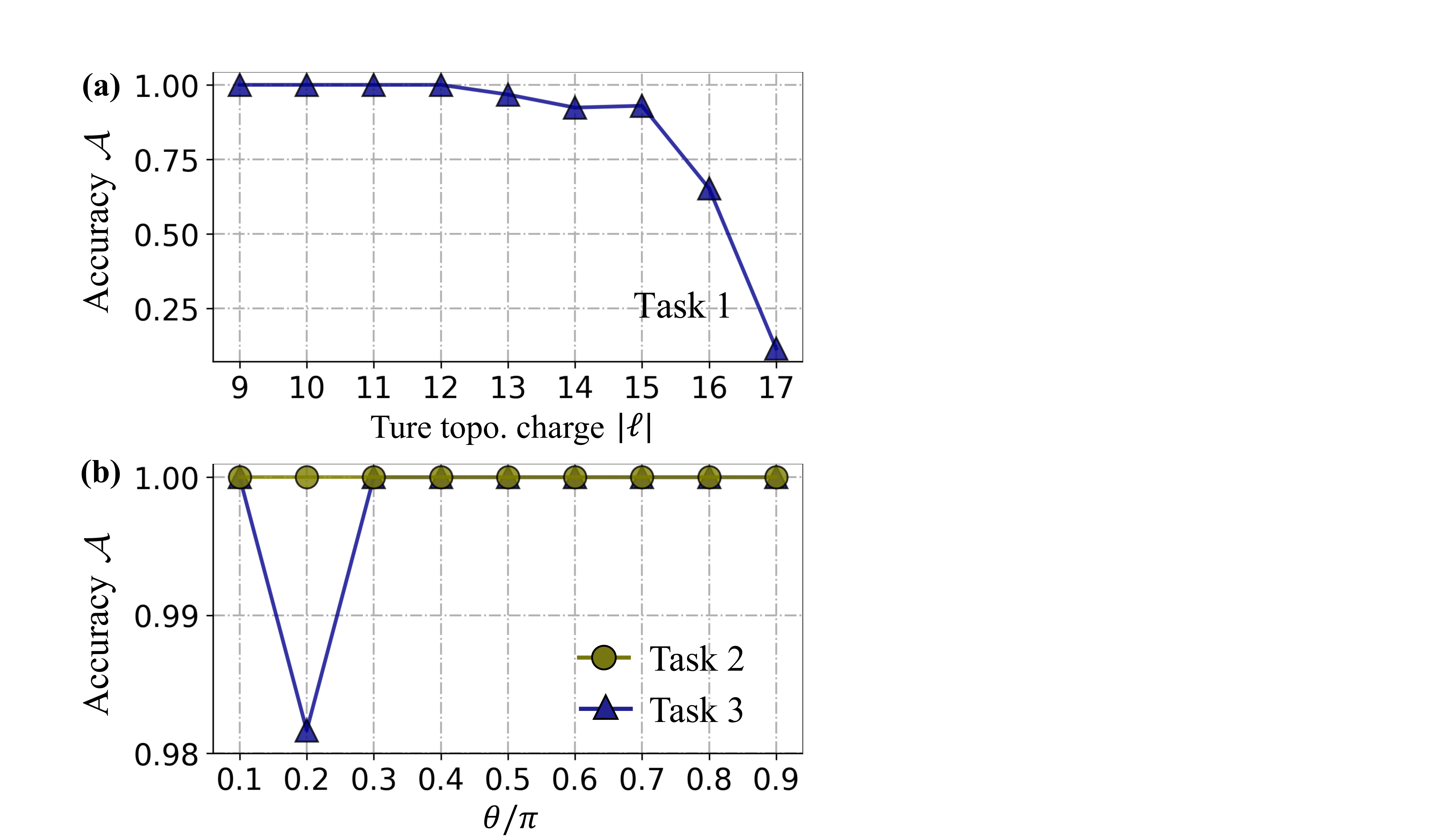}
    \caption{(Color online) Average accuracy $\mathcal{A}$ against (a) $|\ell|$ beyond training areas for Task 1; and (b) $\theta$ for arbitrary superpositions of two OAM modes in Task 2 and Task 3. }
    \label{fig5}
\end{figure}

{\color{blue}\textit{Discussion and conclusion.---}}
Our results have explicitly showcased the exceptional power of machine learning in the experimental recognition of OAMs with large topological charges $\ell$. To further demonstrate the versatility of the CNN, we revisit Task 1 and harness its generalization power to recognize larger $\ell$ beyond training intervals. We consider training intervals $\ell\in \{\pm1,\pm2,...\pm8\}$ ($\ell\in \{\pm1,\pm2,...\pm15\}$ in the SM \cite{SM}) which are accessible in our experiments (simulations). We test the extrapolation ability of the trained CNN by feed 400 images with different rotation phases for each $\ell$. The accuracy for simulated data set against $|\ell|$ is shown in Fig.~\ref{fig5}(a) (Fig. \ref{figS3} in the SM \cite{SM}). Remarkably, we find that the trained CNN can precisely predict topological charges beyond training intervals up to $|\ell|=15$ ($|\ell|=25$).

Furthermore, we show that our approach has great feasibility to classify arbitrary superpositions of two OAM modes $\ket{\psi_{\ell_1,\ell_2}} = \cos\frac{\theta}{2}\ket{\text{LG}_{\ell_1}} + e^{i\phi}\sin\frac{\theta}{2}\ket{\text{LG}_{\ell_2}})$ with various coefficients $\theta$ (and $\phi$) by training few samples. We revisit Task 2 and Task 3 and numerically generate 300 images with $\phi\in[0,2\pi]$ per class for $\theta/\pi=\{0.1,0.5,0.9\}$ and $\{0.1,0.5,0.7,0.9\}$ as the training set, respectively. The test sets consist of 100 images with $\phi\in[0,2\pi]$ for each class $\theta/\pi=\{0.1,0.2,...,0.9\}$, which excludes the data in the training set. After training, we use the test sets to evaluate the performance of the CNN in Task 2 and Task 3. The average accuracy $\mathcal{A}$ against $\theta$ for the two tasks is shown in Fig.~\ref{fig5}(b), which indicates the trained CNN can predict the random topological charges of arbitrary superpositions of two OAM modes with $\mathcal{A}>98\%$.

In summary, we have trained the neutral network to sort OAM modes using their intensity images generalized in simulations and experiments as the input data. We have illustrated that the trained CNN has great generalization power to recognize OAM modes of large topological charges beyond training intervals with nearly 100\% accuracy. We have shown that the CNN can classify arbitrary superpositions of two OAM modes with random topological charges. Our results demonstrate the experimental feasibility of the ML approach to study the OAM physics and to increase the state space of OAM beams in further practical applications.

\begin{acknowledgments}
This work was supported by the National Natural Science
Foundation of China (Grants No. U1830111 and No. 12047522, No. 12074180, and No. U1801661), the Key-Area Research and Development Program of Guangdong Province (Grant No. 2019B030330001), the Science and Technology of Guangzhou (Grants No. 2019050001), and the Guangdong Basic and Applied Basic Research Foundation (Grant No. 2021A1515010315).
\end{acknowledgments}

\bibliography{reference}

\clearpage

\onecolumngrid
\appendix

\section{Supplementary Material for\\Efficient sorting of orbital-angular-momentum states with large topological charges and their unknown superpositions via machine learning}

\subsection{Some details of the CNN and the training}
We present more details of the CNN structure and the training details. Here the CNNs are a specific type of neural network that are generally composed of convolutional, pooling and fully-connected layers. Convolutional layers take advantage of the hierarchical pattern in data and extract features from data efficiently by applying the parameter sharing technique. Pooling layers reduce the redundant information by coarse graining the inputs and thus are useful to extract spatially-invariant features and reduce the weight parameters. Fully-connected layers are used after convolutional and pooling layers to do classification or regression analysis~\cite{LeCun2015}.

\begin{figure}[htb]
    \centering
    \includegraphics[width=0.75\textwidth]{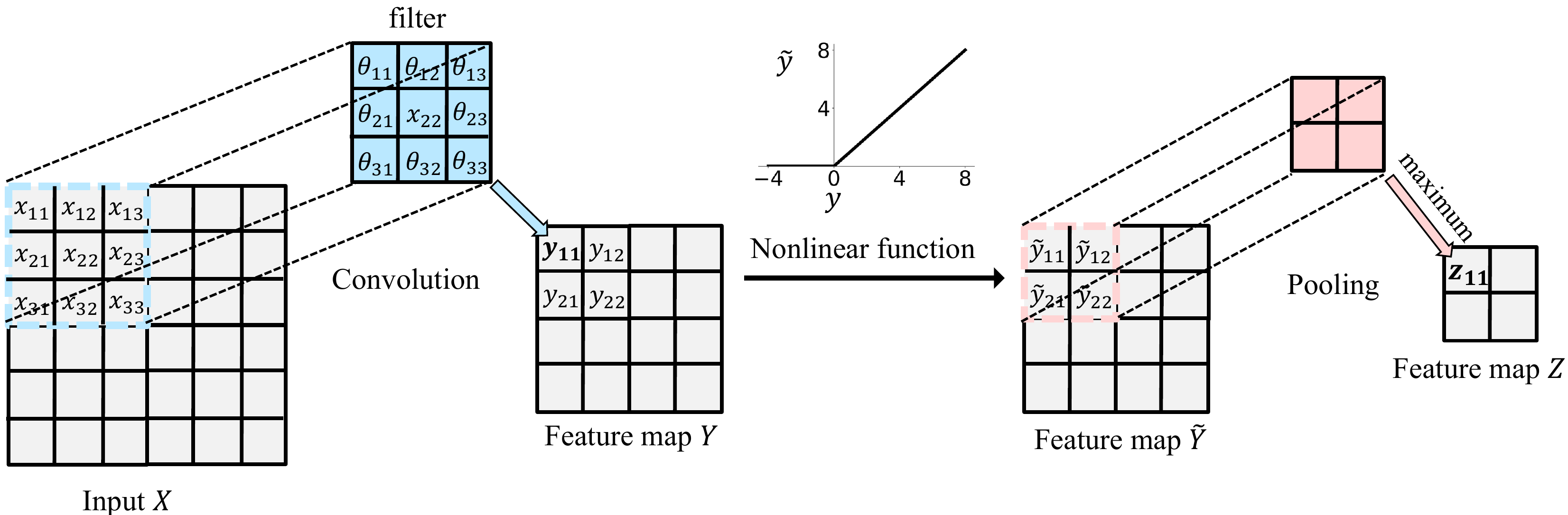}
    \caption{(Color online) The mechanism of the convolutional layer and pooling layer. A convolutional layer operates convolution operations on input $X$. The resulting feature map $Y$ is applied the nonlinear function elementwisely to obtain $\tilde{Y}$. Finally, a max pooling layer downsamples the redundant information by choosing the maximum value in a silding window.
    }
    \label{figS1}
\end{figure}

Figure.~\ref{figS1} shows the mechanism of a convolutional layer and a pooling layer. A convolutional layer use filter that perform convolution operations as it is scanning the input $X$ with respect to its dimensions. Indeed, convolutional layers typically use many filters to extract different features from data. The resulting output $Y$ is called feature map. Following a nonlinear function is applied elementwise to the $Y$. A common choice for the nonlinearity is called Rectified Linear Unit (ReLU), which defined as $x\rightarrow \max{(x,0)}$. The output of the $n$-th convolutional layer can be expressed by
\begin{equation}
    \tilde{Y}_n=f(W_n*X),
\end{equation}
where $f$ denotes the nonlinear activation function, $X$ is the input image or feature map, $W_n$ is the set of filters in the $n$-th convolutional layer and $*$ is the convolution operation. The pooling layer is downsampling operation, typically applied after a convolution layer. Pooling layers coase grain their inputs by partitioning the images into windows of size $k\times k$ and mapping each such window into a single numerical output. A common choice is called max-pooling where the maximum value within the window is taken.
The output of the $n$-th max-pooling layer can be expressed by
\begin{equation}
    Z_{n,ij}=\max_{l,m\in R_{ij}}{\tilde{Y}_{n,lm}}
\end{equation}
where $l,m$ are the locations inside the reception area $R_{i,j}$. Finally, fully-connected layers operate on a flattened features extracted by the convolutional layers and used to classification or regression analysis, depending on the task at
hand.

To construct and train the neural network, we use the deep learning framework PyTorch~\cite{paszke2019pytorch}. We start with a complex multi-layer convolutional network, with a fully connected layer before the output layer. Then we continue to reduce the complexity of neural network, e.g. the number of convolutional filters, the number of layers and the number of neurons in each layer,  as long as it still has good performance. The resulting neural network is reported in the main text. To training the CNNs,  for regression (Task 1), we chose the mean square error as the objective function, which is given by
\begin{equation}\label{eqS4}
    \mathcal{L}_{reg}=\frac{1}{N}\sum_{i=1}^{N}(\tilde{\ell}^{(i)} - \ell^{(i)})^2,
\end{equation}
where $\ell^{(i)}$ ($\tilde{\ell}^{(i)}$) denotes the true (predicted) topological charge of $i$-th image and $N$ is the total number of training data set. For classification analysis (Task 2 and 3), the objective function is the categorical cross-entropy loss which  is defined as
\begin{equation}\label{eqS5}
     \mathcal{L}_{cla}=-\frac{1}{N}\left[\sum_{i=1}^{N}\sum_{j=1}^{n_c}1(l^{(j)}=\tilde{l}^{(j)})\log_2(P_j)\right],
\end{equation}
where $l^{(j)}$ ($\tilde{l}^{(j)}$) is the true(predicted) label of $j$-th image, $P_j$ denotes the output probabilities of $j$-th class, the expression $1(l^{(j)}=\tilde{l}^{(j)})$ means that it will take the value 1 when the condition $l^{(j)}=\tilde{l}^{(j)}$ is statisfied and the value 0 in the opposite case and $n_c$ is the number of class. $n_c=12$ and 6 for Task 2 and 3, respectively. The weight parameters of CNN will be updated in the training process by minimizing Eq. (\ref{eqS4}) or Eq. (\ref{eqS5}), depending on the task at hand. In addition to the training scheme mentioned in the main text, we also use data augmentation by adding slightly modified copies of data during the training to reduce overfitting. Geometric transformations, cropping, rotation, noise injection and random erasing are used to augment image~\cite{Shorten2019}.

\subsection{Sorting OAM modes with larger topological charges}

\begin{figure}[htb]
    \centering
    \includegraphics[width=0.6\textwidth]{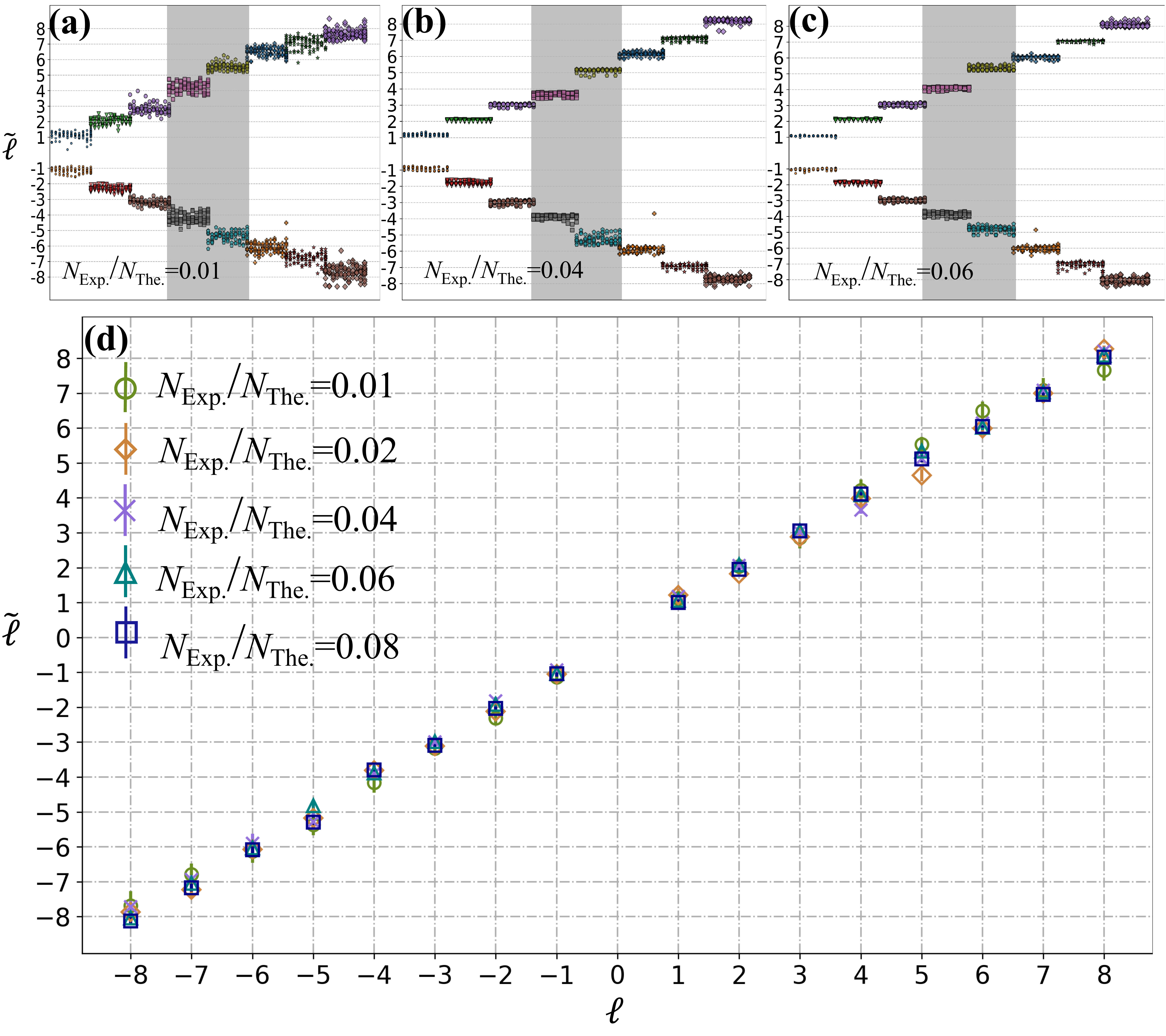}
    \caption{(Color online) (a)-(c) The distribution of output
$\tilde{\ell}$ of convolutional neural network when different fractions of experimental of experimental images are added in the training set for $N_{\text{Exp.}}/N_{\text{The.}}=0.01,0.04,0.06$, respectively. (d) Mean values and error bars of predicted $\tilde{\ell}$ for different $N_{\text{Exp.}}/N_{\text{The.}}$.
    }
    \label{figS2}
\end{figure}

Here we qualitatively show the distribution of output $\tilde{\ell}$ of the CNN when different fractions of experimental images are added in the training set in Figs.~\ref{figS2}(a-c). Clearly, with increasing experimental images added to the training set, the predicted topological charge $\tilde{\ell}$ is more concentrated to the true value $\ell$. Furthermore, we quantitatively show the error bars of predicted topological charges for different fractions $N_{\text{Exp.}}/N_{\text{The.}}$. The mean values of predicted $\tilde{\ell}$ are closer to true $\ell$ and their standard deviations decrease with increasing of $N_{\text{Exp.}}/N_{\text{The.}}$.

\begin{figure}
    \centering
    \includegraphics[width=0.6\textwidth]{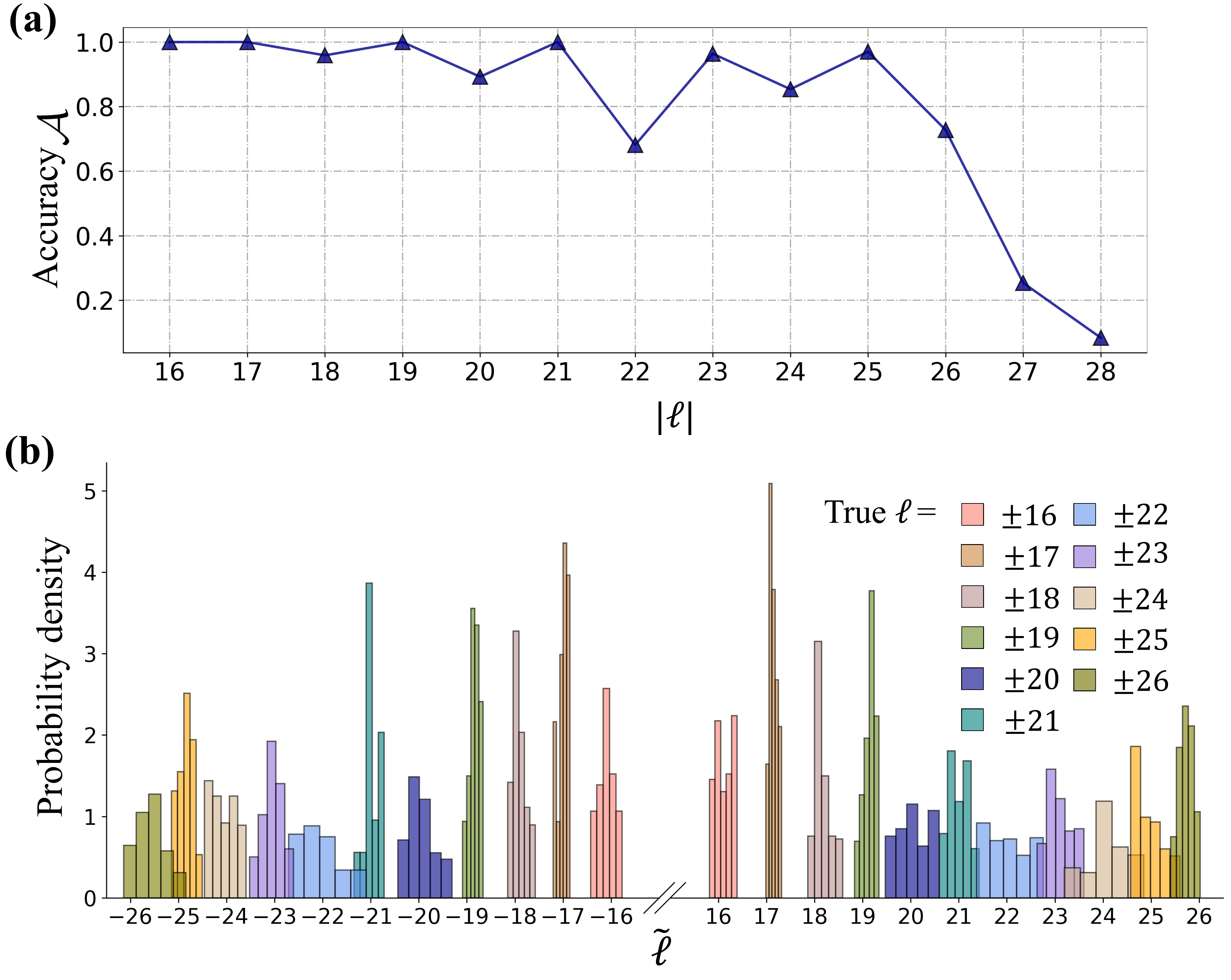}
    \caption{(Color online) (a) The accuracy of the simulated data set synthesized in the intervals beyond training areas against the absolute value of topological charges $|\ell|$. (b) The distribution of output $\tilde{
    \ell}$ including larger intervals $\ell\in\{\pm16,\pm17,...,\pm26\}$ of the CNN trained only in the intervals $\ell\in\{\pm1,\pm2,...,\pm15\}$. The probability for a given $\ell$ sums to unity. There are narrow peaks at each integer.
    }
    \label{figS3}
\end{figure}

To further demonstrate the feasibility of our method in recognizing larger topological charges, we increase the training intervals to $\ell\in\{\pm1, \pm2, ..., \pm15\}$. The accuracy of the simulated data set synthesized in the intervals beyond training areas is shown in Fig.~\ref{figS3}(a). The trained CNN has great generalization power and can recognize topological charge up to $|\ell|=25$ with a high accuracy. The predicted topological charges for all $\ell$s are close to the true values as shown in Fig.~\ref{figS3}(b).

\end{document}